\documentclass
[preprint,pra,showpacs,showkeys,byrevtex,nobibnotes,nofootinbib,12pt,a4paper,onecolumn]{revtex4}%
\usepackage{amsfonts}
\usepackage{amsmath}
\usepackage{amssymb}
\usepackage{graphics}
\usepackage{graphicx}%
\providecommand{\U}[1]{\protect\rule{.1in}{.1in}}

\begin{document}
\title{Non-Hermitian Hamiltonians in decoherence and equilibrium theory}
\author{Mario Castagnino}
\affiliation{CONICET, IAFE (CONICET-UBA), IFIR and FCEN (UBA), Argentina.}
\author{Sebastian Fortin}
\affiliation{CONICET, IAFE (CONICET-UBA) and FCEN (UBA), Argentina.}

\begin{abstract}
There are many formalisms to describe quantum decoherence. However, many of
them give a non general and ad hoc definition of \textquotedblleft pointer
basis\textquotedblright\ or \textquotedblleft moving preferred
basis\textquotedblright, and this fact is a problem for the decoherence
program. In this paper we will consider quantum systems under a general
theoretical framework for decoherence and we will present a tentative
definition of the moving preferred basis. These ideas are implemented in a
well-known open system model. The obtained decoherence and the relaxation
times are defined and compared with those of the literature for the Lee-
Friedrichs model.

\end{abstract}
\keywords{unestable states, complex energy, decoherence, preferred basis, relaxation
time, decoherence time}
\pacs{71.10.Li 03.65.Db 03.65.Yz 03.65.Ta}
\maketitle
\tableofcontents

\section{Introduction}

From the appearance of quantum mechanics many attempts have been made to
recover the laws of the classical mechanics through some classical limit. The
more common scheme of this type includes \textit{quantum decoherence}. This
process eliminates the terms of interference of the density matrix, that are
classically inadmissible, since they prevent the use of a classical (boolean)
logic. In addition, decoherence gives a rule to select candidates for
classical states.

In this work the decoherence is considered an interaction process between an
open quantum system and its environment. This process, called
\textit{Environment-Induced Decoherence} (EID) determines case by case which
is the privileged basis, usually called \textit{moving preferred basis} where
decoherence takes place in a decoherence time $t_{D}$ that is much smaller
than the relaxation time $t_{R}$ and it defines certain observables that
acquire classical characteristics. This is the orthodox position on the
subject \cite{Bub}. The moving preferred basis was introduced, case by case in
several papers (see \cite{Max}) in a non systematic way. On the other hand in
references \cite{OmnesPh} and \cite{OmnesRojo} Roland Omn\`{e}s introduces a
rigorous and almost general definition of the moving preferred basis based in
a reasonable choice of relevant observables, and other physical
considerations. Recently it has become evident that dissipation from system to
environment was not a necessary condition for decoherence \cite{Max} and the
arrival to equilibrium of closed systems was also considered (\cite{GP}%
-\cite{Studies}). Closed system will be discussed at large elsewhere. In this
work we focus our attention on EID, which is a well-known theory, with well
established experimental verifications, which makes unnecessary any further explanation.

Non-unitary evolutions are essential to explain and study decoherence
phenomena, quantum to classical limit, and final equilibrium. These phenomena
appear in the evolution of quantum system, where decoherence time and
relaxation time can be defined using non-unitary evolutions, poles theory, and
non-Hermitian Hamiltonians. We will consider a closed system $U$ and we will
define two subsystems: $S$, the \textquotedblleft proper or open
system\textquotedblright, and $E$, the environment. It is well-known that in
this case the state of the proper system is obtained from the total density
operator by tracing over the environmental degrees of freedom. If we consider
the Hermitian Hamiltonian of a composed closed system $U$ and the inner
product of the evolved state with any observable we can make its analytical
continuation, in the energy variable into the lower complex half-plane, and in
general we will find poles. These poles are complex eigenvalues of the non
Hermitian Hamiltonian $H_{eff}$ that determines the system evolution. These
complex eigenvalues define all the possible non-unitary decaying modes with
characteristic decaying times proportional to the inverse of the imaginary
part of the poles (see \cite{JPA}-\cite{MPLA}). From these characteristic
times we can deduce the relaxation time is the largest characteristic time and
it is related with the pole closest to the real axis. We can also deduce the
decoherence time, that turns out to be a function of the imaginary part of the
poles and the initial conditions of the system. Moreover, we will introduce a
tentative definition of the moving preferred basis. All these definitions are
considered in the Lee Friedrichs model.

\section{\label{GenDefMovDecBas}Towards a definition of the moving preferred
basis.}

In this section we will try to introduce a very general theory for the moving
preferred basis in the case of a general distribution of poles and for
\textit{any relevant observable space} $\mathcal{O}_{R}$. For this purpose it
is necessary to consider the coordinates of observables and states in the
Hamiltonian basis $\{|\omega\rangle\}$ (i.e. the functions $O(\omega
,\omega^{\prime})$ and $\rho(\omega.\omega^{\prime})$) endowed with extra
analytical properties in order to find the definition of a moving preferred basis.

It is well-known that evolution towards equilibrium has two phases (there also
is an initial non exponential Zeno-period which is irrelevant in this paper):

i.- An exponential damping phase that can be described studying the analytical
continuation of the Hamiltonian into the complex plane of the energy (see
\cite{JPA}-\cite{MPLA}).

ii.- A final decaying inverse-polynomial in $t^{-1}$ known as the long time
evolution or Khalfin effect (see \cite{Khalfin}, \cite{BH}), which is
difficult to detect experimentally (see \cite{KhalfinEx}). The power law decay
for long times described by the Khalfin effect has no intrinsic parameter. It
has no characteristic time scale. Khalfin period is the one where the decaying
exponential modes are not more dominant and only inverse powers of time modes
remain. We can consider that the time characteristic of this period is
infinite (or very long). Instead of using the word \textquotedblleft
infinite\textquotedblright, we will use Khalfin time scale.

These two phases will play an important role in the definition of the moving
preferred basis. They can be identified by the theory of analytical
continuation of vectors, observables and states. To introduce the main
equations we will make a short abstract of papers \cite{JPA} and \cite{PRA2}.

\subsection{Analytic continuations in the bra-ket language.}

We begin reviewing the analytical continuation for pure states. Let the
Hamiltonian be $H=H_{0}+V$ where the free Hamiltonian $H_{0}$ satisfies ( see
\cite{JPA} or \cite{PRA2})%

\begin{equation}
H_{0}|\omega\rangle=\omega|\omega\rangle,\text{ }\langle\omega|H_{0}%
=\omega\langle\omega|,\text{ \ \ }0\leq\omega<\infty
\end{equation}
and
\begin{equation}
I=\int_{0}^{\infty}d\omega|\omega\rangle\langle\omega|,\text{ }\langle
\omega|\omega^{\prime}\rangle=\delta(\omega-\omega^{\prime})\label{I}%
\end{equation}
Then
\begin{equation}
H_{0}=\int_{0}^{\infty}\omega|\omega\rangle\langle\omega|d\omega\label{h0}%
\end{equation}
and
\begin{equation}
H=H_{0}+V=\int_{0}^{\infty}\omega|\omega\rangle\langle\omega|d\omega+\int
_{0}^{\infty}d\omega\int_{0}^{\infty}d\omega^{\prime}V_{\omega\omega^{\prime}%
}|\omega\rangle\langle\omega^{\prime}|=\int_{0}^{\infty}\omega|\omega
^{+}\rangle\langle\omega^{+}|d\omega\label{Hamil}%
\end{equation}
where the $|\omega^{+}\rangle$ are the eigenvectors of $H$, that also satisfy
eq. (\ref{I}). The eigen vectors of $H$ are given by the Lippmann-Schwinger
equations (see \cite{JPA}. eq. (12) and (13))%
\begin{equation}
\langle\psi|\omega^{+}\rangle=\langle\psi|\omega\rangle+\langle\psi|\frac
{1}{\omega+i0-H}V|\omega\rangle,\text{ \ \ }\langle\omega^{+}|\varphi
\rangle=\langle\omega|\varphi\rangle+\langle\omega|V\frac{1}{\omega
-i0-H}|\varphi\rangle\label{AN}%
\end{equation}
Let us now endow the function of $\omega$ with adequate analytical properties
(see \cite{Bohm}). E.g. let us consider that the state $|\varphi\rangle$
(resp. $\langle\psi|$) is such that it does not create poles in the complex
extension of $\langle\omega|\varphi\rangle$ (resp. in $\langle\psi
|\omega\rangle$) and therefore this function is analytic in the whole complex
plane. The physical meaning of this hypothesis is that if the system would be
a non interacting one it would never reach equilibrium. Moreover we will
consider that the complex extensions of function $\langle\omega^{+}%
|\varphi\rangle$ (resp. $\langle\psi|\omega^{+}\rangle$) is analytic but with
just one simple pole at $z_{0}$ $=\omega_{0}-\frac{i}{2}\gamma_{0},$
$\gamma_{0}>0$ in the lower halfplane (resp. another pole $z_{0}^{\ast}%
=\omega_{0}+\frac{i}{2}\gamma_{0},\gamma_{0}>0$ on the upper halfplane) (see
\cite{DT} for details). Then in this paper, for the sake of simplicity we will
always use a model with just one pole and an integral that corresponds to the
Khalfin effect. Then we make an analytic continuation from the positive
$\omega$ axis to some curve $\Gamma$ of the lower half-plane.

Then (see \cite{JPA}. eq. (29)) we can define%
\begin{align}
\langle\widetilde{f_{0}}|\varphi\rangle & \equiv cont_{\omega^{\prime
}\rightarrow z_{0}}\langle\omega^{\prime+}|\varphi\rangle,\text{ \ \ }%
\langle\psi|f_{0}\rangle\equiv(-2\pi i)cont_{\omega^{\prime}\rightarrow z_{0}%
}(\omega^{\prime}-z_{0})\langle\psi|\omega^{+}\rangle\nonumber\\
\langle\widetilde{f_{z^{\prime}}}|\varphi\rangle & \equiv cont_{\omega
^{\prime}\rightarrow z^{\prime}}\langle\omega^{\prime+}|\varphi\rangle,\text{
\ \ }\langle\psi|f_{z^{\prime}}\rangle\equiv cont_{\omega^{\prime}\rightarrow
z}\langle\psi|\omega^{+}\rangle,\text{\ }z^{\prime}\in\Gamma,\forall\text{
}|\varphi\rangle\text{ }\langle\psi|\label{L.1}%
\end{align}
and (see \cite{JPA}. eq. (31))%
\begin{align}
\langle\psi|\widetilde{f_{0}}\rangle & \equiv cont_{\omega\rightarrow
z_{0}^{\ast}}\langle\psi|\omega^{+}\rangle,\text{ \ \ }\langle f_{0}%
|\varphi\rangle\equiv(2\pi i)cont_{\omega^{\prime}\rightarrow z_{0}^{\ast}%
}(\omega-z_{0})\langle\omega^{+}|\varphi\rangle\nonumber\\
\langle\psi|\widetilde{f_{z^{\prime}}}\rangle & \equiv cont_{\omega\rightarrow
z}\langle\psi|\omega^{+}\rangle,\text{ \ \ }\langle f_{z}|\varphi\rangle\equiv
cont_{\omega\rightarrow z}\langle\omega^{+}|\varphi\rangle,\text{ \ }%
z\in\Gamma,\forall\text{ }|\varphi\rangle\text{ }\langle\psi|\label{L.2}%
\end{align}
where $cont$ means analytic continuation. The tilde in $\langle\widetilde
{f}_{0}|$ is originated in the fact that \ in the complex extension there is
no one-to-one correspondence between bra and kets \cite{JPA}.

Finally it can be proved that (see \cite{JPA} eq. 1.33 and \cite{PRA2}%
\ eq.82)
\begin{equation}
H=z_{0}|f_{0}\rangle\langle\widetilde{f_{0}}|+\int_{\Gamma}z|f_{z}%
\rangle\langle\widetilde{f_{z}}|dz\label{star}%
\end{equation}
That is a simple extension of the eigen-decomposition of $H$ to the complex
plane with the one-pole term and the integral term that produces the Khalfin effect.

When it is possible to neglect the Khalfin term (i. e. for not extremely long
times) the Hamiltonian reads where we have only a complex energy $z_{0}$.%

\begin{equation}
H_{eff}=z_{0}|f_{0}\rangle\langle\widetilde{f_{0}}|\label{Heff}%
\end{equation}
This is the non Hermitian Hamiltonian that determines the evolution of the
system far from the Khalfin time scale.

\subsection{Analytical continuation in the observables and states language.}

What we have said about the pure states and the Hamiltonian can be rephrased
in the case of the states, observables, and the Liouvillian operator $\ L$
(see a review in \cite{Cos}). But we prefer to follow the line of \cite{JPA}
and keep the Hamiltonian framework and discuss the analytical continuation of
\ $\langle O\rangle_{\rho(t)},$ that we will also symbolize as $(\rho(t)|O)$.
In fact, we know that this scalar is the main character of the play so we will
completely study its analytical properties. So let us call
\begin{equation}
|\omega)=|\omega\rangle\langle\omega|,\text{ and }|\omega,\omega^{\prime
})=|\omega\rangle\langle\omega^{\prime}|\label{A1}%
\end{equation}
Then a generic relevant observable is $O_{R}\in\mathcal{O}_{R}$%
\begin{equation}
O_{R}=|O_{R})=\int d\omega O(\omega)|\omega)+\int d\omega\int d\omega^{\prime
}O(\omega,\omega^{\prime})|\omega,\omega^{\prime})\label{A}%
\end{equation}
and the generic states is%
\begin{equation}
\rho_{R}=(\rho_{R}|=\int d\omega\rho(\omega)\widetilde{(\omega|}+\int
d\omega\int d\omega^{\prime}\rho(\omega,\omega^{\prime})\widetilde
{(\omega,\omega^{\prime}|}\label{B}%
\end{equation}
where (see also \cite{PRA2} eq. (44) or \cite{JPA}. eq. (45)).%
\begin{equation}
\widetilde{(\omega|}\omega^{\prime})=\delta(\omega-\omega^{\prime})\text{
\ \ \ and \ \ \ }\widetilde{(\omega,\omega^{\prime}|}\omega^{\prime\prime
},\omega^{\prime\prime\prime})=\delta(\omega-\omega^{\prime\prime}%
)\delta(\omega^{\prime}-\omega^{\prime\prime\prime})\label{B2}%
\end{equation}
Then%
\begin{equation}
\widetilde{(\omega|}O_{R})=O(\omega),\text{ }\widetilde{(\omega,\omega
^{\prime}|}O_{R})=O(\omega,\omega^{\prime})\label{B3}%
\end{equation}
We will consider the subject as general as possible, i.e. $O_{R}$ would be any
observable such that $O_{R}\in\mathcal{O}_{R}$\ and $\rho_{R}$ any state
$\rho_{R}\in\mathcal{O}_{R}^{\prime}$. In fact, in the next subsection we will
only consider the generic mean value $(\rho_{R}(t)|O_{R}) $ for two
paradigmatic models below. Model 1 with just one pole and the Khalfin effect
and Model 2 with two poles.

\subsection{Model 1. One pole and the Khalfin term:}

We will use a formalism for states and observables which has been proposed by
the Brussels school (led by Ilya Prigogine) in \cite{Brussels}. It can be
proved (cf. (\cite{JPA}) eq. (67)) that the evolution equation of the mean
value $(\rho(t)|O)$ is
\begin{equation}
\langle O_{R}\rangle_{\rho(t)}=(\rho(t)|O_{R})=\int_{0}^{\infty}\rho^{\ast
}(\omega)O(\omega)\,d\omega+\int_{0}^{\infty}\int_{0}^{\infty}\rho^{\ast
}(\omega,\omega^{\prime})O(\omega,\omega^{\prime})\,e^{i\left(  \omega
-\omega^{\prime}\right)  t}\,d\omega d\omega^{\prime}\label{C}%
\end{equation}
i.e. this mean value in the case $V\neq0$ reads%

\begin{equation}
(\rho(t)|O_{R})=\int d\omega(\rho(0)|\Phi_{\omega})(\widetilde{\Phi_{\omega}%
}|O_{R})+\int d\omega\int d\omega^{\prime}e^{i(\omega-\omega^{\prime})t}%
(\rho(0)|\Phi_{\omega\omega^{\prime}})(\widetilde{\Phi_{\omega\omega^{\prime}%
}}|O_{R})
\end{equation}

Where $O(\omega)=(\widetilde{\Phi_{\omega}}|O_{R}),$ $O(\omega,\omega^{\prime
})=(\widetilde{\Phi_{\omega\omega^{\prime}}}|O_{R}),$ $\rho^{\ast}%
(\omega)=(\rho_{R}(0)|\Phi_{\omega}),$ $\rho^{\ast}(\omega,\omega^{\prime
})=(\rho_{R}(0)|\Phi_{\omega\omega^{\prime}})$. These $\Phi$ vectors are
defined as%
\begin{equation}
|\Phi_{\omega})=|\omega^{+}\rangle\langle\omega^{+}|,\text{ }|\Phi
_{\omega\omega^{\prime}})=|\omega^{+}\rangle\langle\omega^{+\prime
}|,\label{53'}%
\end{equation}
and%
\begin{align}
(\widetilde{\Phi_{\omega\omega^{\prime}}}|  & =\int d\varepsilon\lbrack
\langle\omega^{+}|\varepsilon\rangle\langle\varepsilon|\omega^{\prime+}%
\rangle-\delta(\omega-\varepsilon)\delta(\omega^{\prime}-\varepsilon
)](\widetilde{\varepsilon}|+\int d\varepsilon\int d\varepsilon^{\prime}%
\langle\omega^{+}|\varepsilon\rangle\langle\varepsilon^{\prime}|\omega
^{\prime+}\rangle\widetilde{(\varepsilon,\varepsilon^{\prime}}|\nonumber\\
\widetilde{(\Phi_{\omega}|}  & =\widetilde{(\omega|}\label{54}%
\end{align}
It should be emphasized that according to definitions (\ref{A1})-(\ref{B3}),
$\widetilde{(\omega|}\neq\left(  |\omega^{+}\rangle\langle\omega^{+}|\right)
^{\dag}$\ and $\widetilde{(\omega\omega^{\prime}|}\neq\left(  |\omega
^{+}\rangle\langle\omega^{+\prime}|\right)  ^{\dag}$ in contrast to the case
of discrete spectra (see \cite{JPA} for details). Then, if we endow the
functions with analytical properties and there is just one pole $z_{0}$ in the
lower halfplane, we can prove that (see \cite{JPA} eq. (70))%
\begin{align}
(\rho(t)|O_{R})  & =\int d\omega(\rho(0)|\Phi_{\omega})(\widetilde
{\Phi_{\omega}}|O_{R})+e^{i(z_{0}^{\ast}-z_{0})t}(\rho(0)|\Phi_{00}%
)(\widetilde{\Phi_{00}}|O_{R})\nonumber\\
& +\int_{\Gamma}dz^{\prime}e^{i(z_{0}^{\ast}-z^{\prime})t}(\rho(0)|\Phi
_{0z^{\prime}})(\widetilde{\Phi_{0z^{\prime}}}|O_{R})+\int_{\Gamma^{\ast}%
}dze^{i(z-z_{0})t}(\rho(0)|\Phi_{0z})(\widetilde{\Phi_{0z}}|O_{R})\nonumber\\
& +\int_{\Gamma^{\ast}}dz\int_{\Gamma}dz^{\prime}e^{i(z-z^{\prime})t}%
(\rho(0)|\Phi_{zz^{\prime}})(\widetilde{\Phi_{zz^{\prime}}}|O_{R})\label{70}%
\end{align}
where $z_{0}$ is the simple pole in the lower half-plane. $|\Phi
_{z}),(\widetilde{\Phi_{z}}|,|\Phi_{zz^{\prime}}),$ and $(\widetilde
{\Phi_{zz^{\prime}}}|$ can be defined as in the case of eq. (\ref{L.1}) and
(\ref{L.2}). The $|\Phi_{z}),(\widetilde{\Phi_{z}}|,|\Phi_{zz^{\prime}}),$ and
$(\widetilde{\Phi_{zz^{\prime}}}|$ can be also defined as a simple
generalization of the vectors $|f_{0}\rangle,$ $\langle\widetilde{f_{0}}|,$
$|f_{z}\rangle,$ and $\langle\widetilde{f_{z}}|$ (\cite{JPA}. eq. (42)).

Therefore we can conclude than the last four terms of equation (\ref{70})
\ vanish respectively with characteristic times%
\begin{equation}
\frac{1}{\gamma_{0}};\frac{2}{\gamma_{0}};\text{ }\frac{2}{\gamma_{0}}%
;\infty\label{TC}%
\end{equation}
Let us observe that:

\begin{enumerate}
\item[i.] The vanishing of the second, third, and fourth therms of eq.
(\ref{70}) is an \textit{exponential decaying} corresponding to the first
three terms of eq. (\ref{TC}). This will also be the case in more complicated
models with many poles.

\item[ii.] The $\infty$ in eq. (\ref{TC}) means that the evolution of the last
term of this equation corresponds to a polynomial in $t^{-1\text{ }}$, \ i. e.
to the \textit{Khalfin evolution}. This is a very weak effect detected in 2006
\cite{KhalfinEx}. Therefore if there is a finite number of poles and the curve
$\Gamma$, is below them, the contribution of the integral along $\Gamma$
corresponds to the Khalfin effect. A closed system model for Khalfin effect
can be found in \cite{K1}, section 6, and an EID-like model in \cite{K2},
section 5.
\end{enumerate}

Then as we must have $t_{D}\ll t_{R}$ and since from eq. (\ref{TC}) we have
just two characteristic times $\gamma_{0}^{-1}$ and \textquotedblleft$\infty
$\textquotedblright, the only possible choice is $t_{D}=\gamma_{0}^{-1}$\ and
$t_{R}=\infty$. In fact, for times $t\gg$ $t_{D}=\gamma_{0}^{-1}$, eq.
(\ref{70}) reads%
\begin{equation}
(\rho(t)|O_{R})=\int d\omega(\rho(0)|\Phi_{\omega})(\widetilde{\Phi_{\omega}%
}|O_{R})+\int_{\Gamma^{\ast}}dz\int_{\Gamma}dz^{\prime}e^{i(z-z^{\prime}%
)t}(\rho(0)|\Phi_{zz^{\prime}})(\widetilde{\Phi_{zz^{\prime}}}|O_{R}%
)\label{700}%
\end{equation}
since where $t\gg$ $t_{D}=\gamma_{0}^{-1}$ the pole terms have vanished and we
just have the Khalfin term. Let us now diagonalize $\rho(t)$ of the last
equation as%
\begin{equation}
\rho(t)=\int di\rho_{i}(t)|i(t)\rangle\langle i(t)|\label{asterisco}%
\end{equation}
where $\{|i(t)\rangle\}$ is the moving eigenbasis of $\rho(t)$. Now let us
define a state $(\rho_{P}(t)|$, the \textit{preferred state}, such that,
\textit{for all times,} it is%
\begin{equation}
(\rho_{P}(t)|O_{R})=\int d\omega(\rho(0)|\Phi_{\omega})(\widetilde
{\Phi_{\omega}}|O)+\int_{\Gamma^{\ast}}dz\int_{\Gamma}dz^{\prime
}e^{i(z-z^{\prime})t}(\rho(0)|\Phi_{zz^{\prime}})(\widetilde{\Phi_{zz^{\prime
}}}|O)\label{asterisco'}%
\end{equation}
So $\rho_{P}(t)$ is a state that evolves in a model with no poles and with
just the Khalfin term. The functional $(\rho_{P}(t)|$ is defined by the inner
product $(\rho_{P}(t)|O_{R})$ as follows from the Riezs theorem\footnote{All
these formulas are confirmed by the coincidence of results with other methods:
e.g. those used to study a $^{208}Pb(2d_{5/2})$ proton state in a Woods-Saxon
potential (see \cite{JPA} Figure 3).}.

It is quite clear that

\begin{enumerate}
\item[i.] when $t<t_{D}$, $\rho(t)\neq$ $\rho_{P}(t)$

\item[ii.] when $t\rightarrow t_{D}$, $\rho(t)\rightarrow$ $\rho_{P}(t)$

\item[iii.] when $t\gg t_{D}$, $\rho(t)=$ $\rho_{P}(t)$
\end{enumerate}

The eigen states of the $\rho_{P}(t)$ are those that we will choose for the
moving decoherence basis. In fact, diagonalizing $\rho_{P}(t)$ we have
\begin{equation}
\rho_{P}(t)=\sum_{j}\rho_{j}(t)\widetilde{|j(t)\rangle}\widetilde{\langle
j(t)|}\label{asterisco'-b}%
\end{equation}
and when $t\rightarrow$ $t_{D}=\gamma_{0}^{-1}$ we have that $\rho
(t)\rightarrow\rho_{P}(t)$ so from eqs. (\ref{asterisco}) and
(\ref{asterisco'-b}) we see that the eigenbasis of $\rho(t)$ and $\rho_{P}(t)$
also converge
\begin{equation}
\left\{  |i(t)\rangle\right\}  \rightarrow\{\widetilde{|j(t)\rangle
}\}\label{asterisco'-c}%
\end{equation}
Namely the basis $\left\{  |i(t)\rangle\right\}  $ converges to $\{\widetilde
{|j(t)\rangle}\}$ and therefore $\rho(t)$ becomes diagonal in $\{\widetilde
{|j(t)\rangle}\}$. Thus $\{\widetilde{|j(t)\rangle}\}$ is our definition for
the \textit{moving preferred basis} for this case. Since $\rho(t)$ becomes
diagonal in the just defined preferred basis $\{\widetilde{|j(t)\rangle}\}$
when $t\rightarrow t_{D}$ and $t_{D}=\gamma_{0}^{-1}$ is really the definition
of the decoherence time. In this model the relaxation time $t_{R}$ corresponds
with the Khalfin term, i.e. an extremely long time, so that%
\begin{equation}
t_{D}\ll t_{R}\label{nuevo-0}%
\end{equation}

\subsection{Model 2: Two poles without the Khalfin term.}

The Khalfin term is so small (see \cite{KhalfinEx}) that it can be neglected
in most of the experimental cases. Then we can eliminate the Khalfin term
since it corresponds to extremely long time. In this case the Hamiltonian
becomes non-Hermitian as in eqs. (\ref{Heff}) and (\ref{81''}). So let us
consider the case of two poles $z_{0}$ and $z_{1}$ (and no relevant Khalfin
term) where eq. (\ref{70}) reads:%

\begin{align}
(\rho(t)|O_{R})  & =\int d\omega(\rho(0)|\Phi_{\omega})(\widetilde
{\Phi_{\omega}}|O_{R})+e^{i(z_{0}^{\ast}-z_{0})t}(\rho(0)|\Phi_{00}%
)(\widetilde{\Phi_{00}}|O_{R})+\nonumber\\
& +e^{i(z_{1}^{\ast}-z_{0})t}(\rho(0)|\Phi_{10})(\widetilde{\Phi_{10}}%
|O_{R})+e^{i(z_{0}^{\ast}-z_{1})t}(\rho(0)|\Phi_{01})(\widetilde{\Phi_{01}%
}|O_{R})\nonumber\\
& +e^{i(z_{1}^{\ast}-z_{0})t}(\rho(0)|\Phi_{11})(\widetilde{\Phi_{11}}%
|O_{R})\label{70'}%
\end{align}
where $z_{0}$ $=\omega_{0}-\frac{i}{2}\gamma_{0},$ $\gamma_{0}>0$ , $z_{1} $
$=\omega_{1}-\frac{i}{2}\gamma_{1},$ $\gamma_{1}>0,$ and we will also consider
that $\gamma_{0}\ll\gamma_{1}$ (see \cite{Manoloetal} section 3, for details).
Then the characteristic times (\ref{TC}) now read%
\begin{equation}
\frac{1}{\gamma_{0}};\frac{1}{\gamma_{1}+\gamma_{0}}=\frac{1}{\gamma
_{1}+\gamma_{0}}\approx\frac{1}{\gamma_{1}}\label{TC'}%
\end{equation}
Then we must choose $t_{D}=\gamma_{1}^{-1}$\ and $t_{R}=\gamma_{0}^{-1}$. Now
for times $t\gg$ $t_{D}=\gamma_{1}^{-1}$, eq. (\ref{700}) reads%
\begin{equation}
(\rho(t)|O_{R})=\int d\omega(\rho(0)|\Phi_{\omega})(\widetilde{\Phi_{\omega}%
}|O_{R})+e^{i(z_{0}^{\ast}-z_{0})t}(\rho(0)|\Phi_{00})(\widetilde{\Phi_{00}%
}|O_{R})
\end{equation}
and we can define a state $(\rho_{P}(t)|$ such that, it would be%
\begin{equation}
(\rho_{P}(t)|O_{R})=\int d\omega(\rho(0)|\Phi_{\omega})(\widetilde
{\Phi_{\omega}}|O_{R})+e^{i(z_{0}^{\ast}-z_{0})t}(\rho(0)|\Phi_{00}%
)(\widetilde{\Phi_{00}}|O_{R})\label{asterisco''}%
\end{equation}
for \textit{all times}. Repeating the reasoning from eqs. (\ref{700}) to
(\ref{asterisco'-c}) we can see that, diagonalizing this last equation, as in
eq. (\ref{asterisco'-b}), we obtain the moving preferred basis. Then in this
case we see that the \ relaxation is obtained by an exponential damping (not a
Khalfin term) and
\begin{equation}
t_{R}=\frac{1}{\gamma_{0}}\gg t_{D}=\frac{1}{\gamma_{1}}\label{nuevo}%
\end{equation}
Again, in this case when $t\rightarrow$ $t_{D}=\gamma_{0}^{-1}$ we have that
$\rho_{R}(t)\rightarrow\rho_{P}(t)$, and we can conclude that the eigenbasis
of $\rho(t)$ and $\rho_{P}(t)$ also converge as in eq. (\ref{asterisco'-c}).
Namely $\rho(t)$ becomes diagonal in the moving preferred basis in a time
$t_{D}$.

\subsection{The general case}

Let us now consider the general case of a system with $N+1$ poles at
$z_{i}=\omega_{i}^{\prime}-i\gamma_{i}$. In this case it is easy to see that
eq. (\ref{70'}) (with no Khalfin term) becomes:%
\begin{equation}
(\rho(t)|O_{R})=(\rho_{\ast}|O_{R})+\sum_{i=0}^{N}a_{i}(t)\exp\left(
-\gamma_{i}t\right)  =(\rho_{R\ast}|O_{R})+f(t)\label{Suma}%
\end{equation}
where $(\rho_{\ast}|O_{R})$ is the final equilibrium value of $(\rho
(t)|O_{R})$ and the $a_{i}(t)$ are oscillating functions$.$ In the most
general case the $z_{i}$ will be placed either at random or following some
laws. Anyhow in both cases they can be ordered as\footnote{For simplicity we
will only consider the case $\gamma_{0}\ll\gamma_{1}\ll\gamma_{2}\ll...$Other
special cases will be considered elsewhere.}%
\begin{equation}
\gamma_{0}\ll\gamma_{1}\ll\gamma_{2}\ll...
\end{equation}

Then if $\gamma_{0}\ll\gamma_{1}$ it is quite clear that the relaxation time
is
\begin{equation}
t_{R}=\frac{1}{\gamma_{0}}%
\end{equation}
So the relaxation time is defined with no ambiguity. Let us now consider the
decoherence time. Really each pole $z_{i}$ defines a decaying mode with
characteristic time $t_{i}=\gamma_{i}^{-1}$. Moreover the poles contain the
essence of the decaying phenomenon and the definition of the decoherence time
depends on their distribution and other data like the initial conditions. In
fact, the initial conditions seam essential for the definition \ of $t_{D}$.
To introduce these conditions, let us define:
\begin{equation}
f(t)=\sum_{i=0}^{N}a_{i}(t)e^{-\gamma_{i}t},\text{ \ }f^{\prime}(t)=\sum
_{i=0}^{N}a_{i}^{^{\prime}}(t)e^{-\gamma_{i}t}-a_{i}(t)\gamma_{i}%
e^{-\gamma_{i}t}\label{35'}%
\end{equation}
so at $t=0$ we can write the initial conditions as%
\begin{equation}
f(0)=\sum_{i=0}^{N}a_{i}(0),\text{ \ \ \ \ \ \ \ \ \ }f^{\prime}(0)=\sum
_{i=0}^{N}a_{i}^{^{\prime}}(0)-\sum_{i=0}^{N}a_{i}(0)\gamma_{i}%
\end{equation}
Let us call $f(t)=const.\exp g(t)\sim\exp g(t)$, and let us make a Taylor
expansion of $g(t)$ as%
\begin{equation}
g(t)=g(0)+g^{\prime}(0)t+\frac{1}{2}g^{\prime\prime}(0)t^{2}+...\label{app}%
\end{equation}
So let us postulate the reasonable hypothesis that the decoherence time is
$t_{D}\ll t_{R}$. Then, in the period before decoherence that we are
considering , precisely $t<t_{D}\ll t_{R}$, we have $\frac{t}{t_{R}}\ll1.$
With this condition we have the approximation:%
\begin{equation}
g(t)=g(0)+g^{\prime}(0)t\label{tesis}%
\end{equation}
where%
\begin{equation}
g(0)=\log f(0),\text{ \ \ \ }g\prime(0)=\frac{f^{\prime}(0)}{f(0)}%
\end{equation}
These equations contain the initial conditions. Then in this approximation:
\begin{equation}
f(t)=e^{g(0)}e^{tg^{\prime}(0)}=f(0)\exp\left(  \frac{\sum_{i=0}^{N}%
a_{i}^{^{\prime}}}{\sum_{i=0}^{N}a_{i}}t\right)  \exp\left(  -\frac{\sum
_{i=0}^{N}a_{i}\gamma_{i}}{\sum_{i=0}^{N}a_{i}}t\right) \label{Omneslike}%
\end{equation}
So we define%
\begin{equation}
\bar{a}_{i}(t)=f(0)\exp\left(  \frac{\sum_{i=0}^{N}a_{i}^{^{\prime}}}%
{\sum_{i=0}^{N}a_{i}}t\right)  \text{\ \ \ \ \ and\ \ \ \ \ }\gamma
_{eff}=\frac{\sum_{i=0}^{N}a_{i}\gamma_{i}}{\sum_{i=0}^{N}a_{i}}%
\text{\ }\label{eff}%
\end{equation}
And (\ref{Omneslike}) becomes%
\begin{equation}
f(t)=\bar{a}_{i}(t)\exp\left(  -\gamma_{eff}t\right) \label{ft}%
\end{equation}
The decoherence time is
\begin{equation}
t_{D}=\frac{1}{\gamma_{eff}}\label{34'}%
\end{equation}
Then $\gamma_{eff}$ and $t_{D}$ are both functions of the initial conditions.
We will see that this $t_{D}$ coincides with the one of the Omn\`{e}s example
in the next subsection.

Let us now consider the definition of the moving preferred basis. It is clear
that, for the time $t\gg t_{D},$ the modes with characteristic times
$t_{i}<t_{D}$ (i.e. $\gamma_{i}>\gamma_{eff}$), that we will call the
\textit{fast modes}, have become negligible in eq. (\ref{Suma}). Then we can
define the functional ($\rho_{P}(t)|$ as%

\begin{equation}
(\rho_{P}(t)|O_{R})=(\rho_{\ast}|O_{R})+\sum_{i=0}^{M}a_{i}(t)\exp\left(
-\gamma_{i}t\right) \label{EV}%
\end{equation}
where the sum in this equation only contains the $M<N$ poles such that
$\gamma_{i}<\gamma_{eff}$, where the $\gamma_{i}$\ correspond to the
\textit{slow modes}. This is our \textit{adiabatic} choice since we have
selected the slow modes of decaying to define $\rho_{P}(t)$ and rejected the
fast modes. Our adiabatic choice corresponds to keep the slow modes and
disregard the fast ones. Thus, for us the \textit{robust} modes are the
\textit{slow} modes since they are \textquotedblleft the less affected by the
interaction with the environment\textquotedblright, that creates the poles, if
compared with the fast modes, and it is usual to say that these robust modes
are those that define the moving preferred basis. In fact:

i.- If the Hamiltonian would only be $H_{0}$ (cf. eq. (\ref{h0})) there would
not be poles (and this is the usual case in the literature). But the complex
extension of the complete Hamiltonian $H$ (cf. eq. (3)) certainly has poles.
Therefore the poles are created by the \textit{interaction} Hamiltonian $V$.

ii.- Thus the slow modes and the fast ones are defined by these poles, and in
the case we are considering, i.e. EID, the poles are defined by the
\textit{interaction }with the environment.

iii.- Then it is reasonable to call \textit{robust} \ the slow modes, since
the environment interaction has smaller influence in these poles, and we
conclude that these are the modes that define the moving preferred basis.

This is our definition of \textit{robustness. \ }Analogously, if we compute
the \textit{linear entropy} we will have a slower variation of this entropy,
if we only consider the slow modes, than if we consider all the modes
(including the fast ones). This would be our minimization of the linear
entropy: the moving preferred basis evolution only contains the slow modes.

Moreover, when $t\gg t_{D}$ the motions produced by the fast modes, such that
$\gamma_{i}>\gamma_{eff},$ \ namely those with motions faster than the one of
the evolution of eq. (\ref{eff}), are no more relevant for $\rho(t)$, and
$\rho_{P}(t)\rightarrow\rho(t)$. Then we diagonalize $\rho_{P}(t)$ and we
obtain the moving preferred basis $\{\widetilde{|j(t)\rangle}\}$. The only
influence in the evolution of $\rho_{P}(t)$ is given the poles such that
$\gamma_{i}<\gamma_{eff}$. When $t\rightarrow t_{D},$ $\left\{  |i(t)\rangle
\right\}  \rightarrow\{\widetilde{|j(t)\rangle}\}$ the eigenbasis of $\rho(t)$
where $0\leq t\leq\infty$. This $\{\widetilde{|j(t)\rangle}\}$ is our
candidate for a general definition of moving preferred basis.

\section{\label{Polos}The Omn\`{e}s or Lee-Friedrichs model.}

Our more complete and simplest example of decoherence in open systems is the
Omn\`{e}s \textquotedblleft pendulum\textquotedblright\ (i. e. oscillator
\cite{Omnesazul}) in a bath of oscillators, that we will compare with the
poles theory in the following subsections. In fact the Omn\`{e}s model could
be considered a poles model if we retain the poles and neglect the Khalfin
term. Moreover in the Omn\`{e}s philosophy the moving preferred basis must be
related to some \textquotedblleft collective variables\textquotedblright\ in
such a way that they would be experimentally accessible. In this case this
variable is the center of mass of the pendulum, i. e. the mean value of the
position of a coherent state. In \cite{Omnesazul} page 285 a one dimensional
\textquotedblleft pendulum\textquotedblright\ (the system) in a bath of
oscillators (the environment) is considered. Then the Hamiltonian reads%
\begin{equation}
H=\omega a^{\dagger}a+\sum_{k}\omega_{k}b_{k}^{\dagger}b_{k}+\sum_{k}%
(\lambda_{k}a^{\dagger}b_{k}+\lambda_{k}^{\ast}ab_{k}^{\dagger})\label{1}%
\end{equation}
where $a^{\dagger}(a)$ is the creation (annihilation) operator for the system,
$b_{k}^{\dagger}(b_{k})$ are the creation (annihilation) operators for each
mode of the environment, $\omega$ and $\omega_{k}$ are the energies of the
system and of each mode of the environment and $\lambda_{k}$ are the
interaction coefficients.

Then let us consider a state%
\begin{equation}
|\psi(t)\rangle=a|\alpha_{1}(t)\rangle\prod_{k}|\beta_{k1}(t)\rangle
+b||\alpha_{2}(t)\rangle\prod_{k}|\beta_{k2}(t)\rangle
\end{equation}
where $|\alpha_{1}(0)\rangle,|\alpha_{2}(0)\rangle$ are \textit{coherent}
states for the \textquotedblleft system\textquotedblright\ corresponding to
the operator $a^{\dagger}$, with center in $x_{1}(0)$\ and $x_{2}%
(0)$\ respectively, and $|\beta_{k1}(0)\rangle,$ $|\beta_{k2}(0)\rangle$ are a
coherent state for the environment corresponding to the operator
$b_{k}^{\dagger}.$ Let the initial conditions be%
\begin{equation}
|\psi(0)\rangle=a|\alpha_{1}(0)\text{ }\{\beta_{k1}(0)=0\}\rangle+b|\alpha
_{2}(t),\{\beta_{k2}(0)=0\rangle\label{in}%
\end{equation}
Moreover Omn\`{e}s shows that, under reasonable hypotheses and approximations
the \textit{relaxation time} of the system is
\begin{equation}
t_{R}=1/\gamma\label{tROmnes}%
\end{equation}
where%

\begin{equation}
\gamma=\pi\int n(\upsilon^{\prime})d\upsilon^{\prime}\lambda_{\upsilon
^{\prime}}^{2}\delta(\omega-\upsilon^{\prime})\label{RT-01}%
\end{equation}
where $n(\upsilon^{\prime})d\upsilon^{\prime}=d\mathbf{k}$. On the other hand,
the decoherence time of the system is (see \cite{Omnesazul}, pp. 289-291)%

\begin{equation}
t_{D}\sim\frac{1}{m\omega_{0}L_{0}^{2}}t_{R}\label{4-0-01}%
\end{equation}
where $L_{0}=$%
$\vert$%
$x_{1}(0)-x_{2}(0)|$. In the next subsection, we will attempt to recover these
results using the polar technique.

\subsection{\label{PolosH}The characteristic times from the polar technique.}

A particular important model can be studied, like the one in \cite{PRE}, with
the Hamiltonian%

\begin{equation}
H=\omega_{0}a^{\dagger}a+\int\omega_{\mathbf{k}}b_{\mathbf{k}}^{\dagger
}b_{\mathbf{k}}d\mathbf{k}+\int\lambda_{\mathit{k}}(a^{\dagger}b_{\mathbf{k}%
}+b_{\mathbf{k}}^{\dagger}a)d\mathbf{k}\label{PRE}%
\end{equation}

i.e. a continuous version of (\ref{1}). In this continuous version we are
forced to endow the scalar $(\rho(t)|O_{R})$ with some analyticity conditions.
Precisely function $\lambda_{\mathit{k}}$ (where $k=\omega_{k}=|\mathbf{k|)}$
is chosen in such a way that%
\begin{equation}
\eta_{\pm}(\omega_{k})=\omega_{k}-\omega_{0}-\int\frac{d\mathbf{k}%
\lambda_{\mathit{k}}^{2}}{\omega_{k}-\omega_{k^{\prime}}\pm i0}\label{cuac-00}%
\end{equation}
does not vanish when $k\in\mathbb{R}^{+},$ and its analytic extension
$\eta_{+}(z)$ in the lower half plane only has a simple pole at $z_{0}$. This
fact will have influence on the poles of $(\rho(t)|O_{R})$ as in the last
section and we know that the study of $(\rho(t)|O_{R})$ is the essential way
to understand the whole problem.

The Hamiltonian (\ref{PRE}) is sometimes called the Lee-Friedrichs Hamiltonian
and it is characterized by the fact that it contains different \textit{number
of modes sector} (number of particle sectors in QFT). In fact, $a^{\dagger}$
and $b_{\mathbf{k}}^{\dagger}$ are creation operators that allow to define
these numbers of mode sectors. e. g. the one mode sector will contain states
like $a^{\dagger}|0\rangle$ and $b_{\mathbf{k}}^{\dagger}|0\rangle$ (where
$a|0\rangle=$ $b_{\mathbf{k}}|0\rangle=0$). Then the action of $\exp\left(
-Ht\right)  $ (or simple the one of $H$) will conserve the number of modes of
this sector in just one mode, since in \ (\ref{PRE}) all the annihilation
operators are preceded by a creation operator. This is also the case for the
$n-$mode sector.

\subsubsection{The Friedrichs model and the relaxation time}

In the case of the one mode sector this model is the so called
Friedrichs-Fano-Anderson or Friedrichs model. For a complete discussion on
this model see \cite{Longhi}. The Hamiltonian of the Friedrichs model is%
\begin{equation}
H_{F}=\omega_{0}\left\vert 1\right\rangle \left\langle 1\right\vert
+\int\omega_{k}\left\vert \omega\right\rangle \left\langle \omega\right\vert
d\omega+\int\left(  \lambda(\omega)\left\vert \omega\right\rangle \left\langle
1\right\vert +\lambda^{\ast}(\omega)\left\vert 1\right\rangle \left\langle
\omega\right\vert \right)  d\omega\label{cuac-04}%
\end{equation}
(this Hamiltonian which is similar to the one of eq. (\ref{Hamil}), is
expressed just in the variable $\omega$, the one that will be analytically
continued). As a consequence of the analyticity condition, mentioned above,
this simple Friedrichs model only shows one resonance. In fact, this resonance
is produced in $z_{0}$. In paper \cite{Longhi} we can see that the poles we
compute here are the same as the poles of the Green%
\'{}%
s function. Let $H_{F}$ be the Hamiltonian of the complex extended Friedrichs
model, i.e. the Hamiltonian of eq. (\ref{star}), then\footnote{Only
symbolically, since the poles really belong to the scalar $(\rho(t)|O)$, as in
the last section.} :%
\begin{equation}
H_{F}|z_{0}\rangle=z_{0}|z_{0}\rangle,\text{ \ \ \ }H_{F}|z\rangle
=z|z\rangle\label{cuac-05}%
\end{equation}
where $z_{0}=\omega_{0}+\delta\omega_{o}-i\gamma_{0}=\omega_{0}^{\prime
}-i\gamma_{0}$ is the only pole and $z\in\Gamma$ corresponds to the integral
term and to the Khalfin effect.

The Lee-Friedrichs model, describing the interaction between a quantum
oscillator and a scalar field, is extensively analyzed in the literature.
Generally, this model is studied by analyzing the one excited mode sector,
i.e. the Friedrichs model. Then, if we compute the pole, of this last model,
up to the second order in $\lambda_{k}$ we obtain that
\begin{equation}
z_{0}=\omega_{0}+\int\frac{d\mathbf{k}^{\prime}\lambda_{\mathit{k}^{\prime}%
}^{2}}{\omega_{0}-\omega_{k}+i0}\label{cuac-01}%
\end{equation}

So the pole (that will corresponds to the one closest to the real axis in the
Lee-Friedrichs model) can be computed (see \cite{LCIB} eq. (42)). These
results coincide (mutatis mutandis) with the one of Omn\`{e}s book
\cite{Omnesazul}\ page 288, for the relaxation time. In fact:
\begin{equation}
\frac{1}{\omega_{0}-\omega^{\prime}+i0}=P\left(  \frac{1}{\omega_{0}%
-\omega^{\prime}}\right)  -i\pi\delta(\omega_{0}-\omega^{\prime}%
)\label{cuac-02}%
\end{equation}
where $P$ symbolizes the \textquotedblleft principal part\textquotedblright,
so%
\begin{equation}
z_{0}=\omega_{0}+P\int\frac{d\mathbf{k}^{\prime}\lambda_{\mathbf{k}^{\prime}%
}^{2}}{\omega_{0}-\omega_{k}}-i\pi\int d\mathbf{k}^{\prime}\lambda
_{\mathbf{k}^{\prime}}^{2}\delta(\omega_{0}-\omega_{k})\label{cuac-03}%
\end{equation}
Then if $d\mathbf{k}=n(\omega)d\omega$ we have%
\begin{equation}
\delta\omega=P\int\frac{n(\omega^{\prime})d\omega^{\prime}\lambda
_{\omega^{\prime}}^{2}}{\omega_{0}-\omega^{\prime}},\text{ \ \ \ \ }\gamma
=\pi\int n(\omega^{\prime})d\omega^{\prime}\lambda_{\omega^{\prime}}^{2}%
\delta(\omega_{0}-\omega^{\prime})\label{Omnes}%
\end{equation}
where $\delta\omega$ is a shift and $\gamma$\ a damping coefficient, then the
system would arrive to a state of equilibrium, namely the results of
\cite{Omnesazul}\ page 288, and the one contained in eq. (\ref{RT-01}) yields:%
\begin{equation}
z_{0}=\left(  \omega_{0}+\delta\omega\right)  -i\gamma=\omega_{0}^{\prime
}-i\gamma\label{cuac}%
\end{equation}
So the Omn\`{e}s result for the relaxation time \textit{coincides}, as we have
already said, with the one obtained by the pole theory, precisely (see
(\ref{tROmnes}))
\begin{equation}
t_{R}=\frac{1}{\gamma}%
\end{equation}

\subsubsection{\label{PolosH copy(2)}Other poles of the Lee-Friedrichs model.}

Let us now consider the Lee-Friedrichs Hamiltonian (\ref{PRE}) for the many
modes sectors. Then, as an example for the three mode sector (with just the
unique pole $z_{0}$ and $z_{1}$, $z_{2}$, or $z_{3}$ \textquotedblleft real
continuous eigenvalues\textquotedblright\ transported to the curve $\Gamma$)
we have:%
\begin{equation}
H|z_{a},z_{b},z_{c}\rangle=(z_{a}+z_{b}+z_{c})|z_{a},z_{b},z_{c}%
\rangle\label{cuac-06}%
\end{equation}
where $(z_{a}+z_{b}+z_{c})$ is the eigenvalue. Then $z_{1},z_{2},z_{3}%
\in\Gamma$ is the Khalfin terms (i.e. they belong to the complex contour on
the lower complex energy plane), and let $z_{0}$ be the pole of one particle
sector. So in the real complex plane the spectrum of $H$ contains

1.- Eigenvalues $(z_{1}+z_{2}+z_{3})$ with three points of the curve $\Gamma$.

2.- Eigenvalues $(z_{1}+z_{2}+z_{0}),$ $(z_{1}+z_{0}+z_{3})$ and $(z_{0}%
+z_{2}+z_{3})$, with two points of the curve $\Gamma$ and the pole $z_{0}$.

3.- Eigenvalues $(z_{1}+z_{0}+z_{0}),$ $(z_{0}+z_{2}+z_{0})$ and $(z_{0}%
+z_{0}+z_{3})$, with a pole at $2z_{0}$, and one point of the curve $\Gamma$.

4.- Eigenvalue $(z_{0}+z_{0}+z_{0})$, with a pole at $3z_{0}$.

These values appear in expression of the mean value as $\sim e^{-i\frac{z_{0}%
}{\hbar}t}$ (like in eq. (19) second term of the l.h.s.) or as $\sim
\int_{\Gamma}e^{-i\frac{z}{\hbar}t}f(z)dz$ (like in eq. (19) three last terms
of the l.h.s.). Then we have that the four cases above become:

1.- $\int_{\Gamma}\int_{\Gamma}\int_{\Gamma}e^{-i\frac{(z_{1}+z_{2}+z_{3}%
)}{\hbar}t}f(z_{1},z_{2},z_{3})dz_{1}dz_{2}dz_{3}$

2.- $\int_{\Gamma}\int_{\Gamma}e^{-i\frac{(z_{1}+z_{2}+z_{0})}{\hbar}t}%
f(z_{1},z_{2})dz_{1}dz_{2}$, and the same for the combinations $(z_{1}%
+z_{0}+z_{3}),$ $(z_{0}+z_{2}+z_{3})$

3.- $\int_{\Gamma}e^{-i\frac{(z_{1}+2z_{0})}{\hbar}t}f(z_{1})dz_{1}$, and the
same for the combinations $(z_{2}-2z_{0}),$ $(z_{3}+2z_{0})$

4.- $e^{-i\frac{3z_{0}}{\hbar}t}$

Then if we neglect the Khalfin we just have the point 4.

Of course in the general case $3\rightarrow n$ we would have $e^{-i\frac
{nz_{0}}{\hbar}t}$ (for the point $n$) , plus many integrals on the curve
$\Gamma$ (for the points \ $1,2,...n-1)$ corresponding to Khalfin terms \ Then
if we neglect the integrals that produce the Khalfin effect, since this effect
corresponds to extremely long times, the $\Gamma$ term disappears and we
simply have a pole at $z_{n}=nz_{0}$. This elimination (in the case of just
one pole $z_{0}$) introduces in the model a structure of a complex oscillator.
Then we can introduce a non Hermitian effective Hamiltonian%
\begin{equation}
H_{eff}=z_{0}\left(  a_{0}^{\dagger}a_{0}+\frac{1}{2}\right)  =z_{0}\left(
N_{0}+\frac{1}{2}\right) \label{81'}%
\end{equation}
where $a_{0}^{\dagger}$ and $a_{0}$ are creation and annihilation operators
and $N_{0}=$ $a_{0}^{\dagger}$ $a_{0}$ is the number of modes operator and%
\begin{equation}
N_{0}|n\rangle=n|n\rangle
\end{equation}
In the case of large $n$, $H_{eff}$\ becomes extremely close to%
\begin{equation}
H_{eff}=z_{0}a_{0}^{\dagger}a_{0}=z_{0}N_{0}\label{81''}%
\end{equation}
Moreover we can call
\begin{equation}
z_{n}=nz_{0}=n(\omega_{0}-i\gamma_{0})\label{cuac-15}%
\end{equation}
and we will find the evolutions%
\begin{equation}
\exp(-iH_{eff}t)|n\rangle=\exp(-inz_{0}t)|n\rangle=\exp(-iz_{n}t)|n\rangle
\end{equation}
\ So, in this approximation, the effective Lee-Friedrichs Hamiltonian
$H_{eff}$ simply is a (non Hermitian) version of $H$ with just damping terms.
We will below use this structure.

\subsubsection{\label{PolosI}The initial conditions}

As an initial conditions, $|\alpha_{1}(0)\rangle,$ $|\alpha_{2}(0)\rangle,$ it
is possible to choose any linear combination of the elements $\left\{
\left\vert n\right\rangle \right\}  $\ where $n=0,1...,\infty$. So we can
choose coherent states
\begin{equation}
\left\vert \alpha_{i}(0)\right\rangle =e^{-\frac{\left\vert \alpha
_{i}(0)\right\vert ^{2}}{2}}\sum_{n=0}^{\infty}\frac{\left(  \alpha
_{i}(0)\right)  ^{n}}{\sqrt{n!}}\left\vert n\right\rangle \label{CI-09c}%
\end{equation}
Then let us choose the initial conditions as the sum of two coherent states,
namely:%
\begin{equation}
\left\vert \Phi(0)\right\rangle =a\left\vert \alpha_{1}(0)\right\rangle
+b\left\vert \alpha_{2}(0)\right\rangle \label{CI-10}%
\end{equation}
Thus the initial state operator is:%
\begin{equation}
\rho(0)=\left\vert a\right\vert ^{2}\left\vert \alpha_{1}(0)\right\rangle
\left\langle \alpha_{1}(0)\right\vert +\left\vert b\right\vert ^{2}\left\vert
\alpha_{2}(0)\right\rangle \left\langle \alpha_{2}(0)\right\vert +ab^{\ast
}\left\vert \alpha_{1}(0)\right\rangle \left\langle \alpha_{2}(0)\right\vert
+a^{\ast}b\left\vert \alpha_{2}(0)\right\rangle \left\langle \alpha
_{1}(0)\right\vert \label{CI-13}%
\end{equation}
We choose the two Gaussian (\ref{CI-09c}) with center at $p_{1,2}(0)=0$, (see
\cite{Omnesazul} eq. (7.15) page 284) and%
\begin{equation}
\alpha_{1}(0)=\frac{m\omega}{\sqrt{2m\omega}}x_{1}(0)\text{ \ \ \ \ and
\ \ \ \ }\alpha_{2}(0)=\frac{m\omega}{\sqrt{2m\omega}}x_{2}(0)\label{CI-17}%
\end{equation}
So $\alpha_{1}(0)$ and $\alpha_{2}(0)$ are real numbers. With a change of
coordinates we can choose $x_{1}(0)$ and $x_{2}(0)$ without loss of
generality. So we can consider that the $\alpha_{1}(0)$ and $\alpha_{2}(0)$
are both positive. For this reason we will interchange $\alpha_{i}(0)$ and
$\left\vert \alpha_{i}(0)\right\vert $ below. With no lost of generality we
can choose
\begin{equation}
\alpha_{1}(0)=0\text{ \ \ \ \ and \ \ \ \ }\alpha_{2}(0)=\frac{m\omega}%
{\sqrt{2m\omega}}L_{0}%
\end{equation}

\emph{The macroscopic case}

It is easy to prove that for macroscopic initial conditions, i.e. when the
peaks of the two Gaussians are far from each other, that is to say
$\vert$%
$\alpha_{1}(0)-\alpha_{2}(0)|\rightarrow\infty$, the states $\left\{
\left\vert \alpha_{1}(0)\right\rangle ,\left\vert \alpha_{2}(0)\right\rangle
\right\}  $\ are quasi-orthogonal basis%
\begin{equation}
\left\langle \alpha_{1}(0)|\alpha_{2}(0)\right\rangle \cong\left\langle
\alpha_{2}(0)|\alpha_{1}(0)\right\rangle \cong e^{-\frac{\left(  \alpha
_{1}(0)-\alpha_{2}(0)\right)  ^{2}}{2}}\cong0\label{CI-21}%
\end{equation}
Then, the macroscopic condition for the initial conditions is $\left\vert
\alpha_{1}(0)-\alpha_{2}(0)\right\vert \gg1$. So we have $\left\vert
\alpha_{1}(0)-\alpha_{2}(0)\right\vert =\alpha_{2}(0)\gg1$,
\begin{equation}
\frac{m\omega}{\sqrt{2m\omega}}L_{0}\gg1
\end{equation}

\subsubsection{Components of the non-diagonal part of the state}

Therefore the evolved state is
\begin{equation}
\rho(t)=\left\vert a\right\vert ^{2}\left\vert \alpha_{1}(t)\right\rangle
\left\langle \alpha_{1}(t)\right\vert +\left\vert b\right\vert ^{2}\left\vert
\alpha_{2}(t)\right\rangle \left\langle \alpha_{2}(t)\right\vert +ab^{\ast
}\left\vert \alpha_{1}(t)\right\rangle \left\langle \alpha_{2}(t)\right\vert
+a^{\ast}b\left\vert \alpha_{2}(t)\right\rangle \left\langle \alpha
_{1}(t)\right\vert \label{CI-16}%
\end{equation}
Let us not consider the non-diagonal part of $\rho(t)$, $\rho^{(ND)}(t)$ in
the basis of the initial conditions $\left\{  \left\vert \alpha_{1}%
(0)\right\rangle ,\left\vert \alpha_{2}(0)\right\rangle \right\}  $. Then we
have%
\begin{equation}
\rho^{(ND)}(t)=\rho_{12}^{(ND)}(t)\left\vert \alpha_{1}(0)\right\rangle
\left\langle \alpha_{2}(0)\right\vert +\rho_{21}^{(ND)}(t)\left\vert
\alpha_{2}(0)\right\rangle \left\langle \alpha_{1}(0)\right\vert \label{CI-20}%
\end{equation}
Since the basis $\left\{  \left\vert \alpha_{1}(0)\right\rangle ,\left\vert
\alpha_{2}(0)\right\rangle \right\}  $ is quasi-orthogonal, from eq.
(\ref{CI-16}) we have
\begin{equation}
\rho_{ij}^{(ND)}(t)=ab^{\ast}\left\langle \alpha_{i}(0)|\alpha_{1}%
(t)\right\rangle \left\langle \alpha_{2}(t)|\alpha_{j}(0)\right\rangle
+a^{\ast}b\left\langle \alpha_{i}(0)|\alpha_{2}(t)\right\rangle \left\langle
\alpha_{1}(t)|\alpha_{j}(0)\right\rangle \label{CI-22}%
\end{equation}
If we consider the evolution given by the non Hermitian Hamiltonian%
\begin{equation}
\left\vert \alpha_{i}(t)\right\rangle =e^{-iH_{eff}t}\left\vert \alpha
_{i}(0)\right\rangle =e^{-\frac{\left\vert \alpha_{i}(0)\right\vert ^{2}}{2}%
}\sum_{n=0}^{\infty}\frac{\left(  \alpha_{i}(0)\right)  ^{n}}{\sqrt{n!}%
}e^{-iz_{n}t}\left\vert n\right\rangle
\end{equation}
we can compute these products $\left\langle \alpha_{i}(0)|\alpha
_{j}(t)\right\rangle $ and we can replace them in (\ref{CI-22}) to obtain%
\begin{align}
\rho_{12}^{(ND)}(t) &  \cong ab^{\ast}e^{-\left\vert \alpha_{2}(0)\right\vert
^{2}\left(  1-e^{iz_{0}^{\ast}t}\right)  }\nonumber\\
\rho_{21}^{(ND)}(t) &  \cong a^{\ast}be^{-\left\vert \alpha_{2}(0)\right\vert
^{2}\left(  1-e^{-iz_{0}t}\right)  }\label{CI-39}%
\end{align}

\subsubsection{Decoherence time}

Since the contributions $\gamma_{n}$ of individual poles $z_{n}$ do not appear
explicitly in the equation (\ref{CI-39}), we may think that such poles are not
involved in the outcome.

However, if we express the exponential of (\ref{CI-39}) as its Taylor series,
we have%
\begin{align}
\rho_{12}^{(ND)}(t) &  \cong ab^{\ast}e^{-\left\vert \alpha_{2}(0)\right\vert
^{2}}\sum_{n=0}^{\infty}\frac{\left(  \left\vert \alpha_{2}(0)\right\vert
^{2}\right)  ^{n}}{n!}e^{iz_{n}^{\ast}t}=ab^{\ast}\sum_{n=0}^{\infty}%
c_{n}(t)e^{-\gamma_{n}t}\nonumber\\
\rho_{21}^{(ND)}(t) &  \cong a^{\ast}be^{-\left\vert \alpha_{2}(0)\right\vert
^{2}}\sum_{n=0}^{\infty}\frac{\left(  \left\vert \alpha_{2}(0)\right\vert
^{2}\right)  ^{n}}{n!}e^{-iz_{n}t}=a^{\ast}b\sum_{n=0}^{\infty}c_{n}^{\ast
}(t)e^{-\gamma_{n}t}%
\end{align}
where%

\begin{equation}
c_{n}(t)=e^{-\left\vert \alpha_{2}(0)\right\vert ^{2}}\frac{\left(  \left\vert
\alpha_{2}(0)\right\vert ^{2}\right)  ^{n}}{n!}e^{i\omega_{n}t}\label{CI-39'}%
\end{equation}
and these equations show that all the diagonal terms vanish when
$t\rightarrow\infty$ showing that there is decoherence. Now we would like to
know the decoherence time, then we must find $\gamma_{eff}$. So we analyze the
decay of
\begin{equation}
\left\vert \rho_{12}^{(ND)}(t)\right\vert ^{2}=\rho_{12}^{(ND)}(t)\left(
\rho_{12}^{(ND)}(t)\right)  ^{\ast}=\left\vert ab\right\vert ^{2}%
e^{g(t)}\label{taylor-01}%
\end{equation}
where%
\begin{equation}
g(t)=\ln\left(  \left(  \sum_{n=0}^{\infty}c_{n}(t)e^{-\gamma_{n}t}\right)
\left(  \sum_{j=0}^{\infty}c_{n}^{\ast}(t)e^{-\gamma_{n}t}\right)  \right)
\label{taylor-02}%
\end{equation}
Let us now expand $e^{g(t)}$ as:
\begin{equation}
e^{g(t)}=e^{g(0)+g^{\prime}(0)t+\frac{1}{2}g^{\prime\prime}(0)t^{2}%
+...}\label{taylor-03}%
\end{equation}
As the decoherence time is a very short one $t_{D}\ll t_{R}$ let us neglect it
from the quadratic term. Now, from eq. (\ref{taylor-02}) we have that%
\begin{align}
g(0) &  =0\nonumber\\
g^{\prime}(0) &  =-2e^{-\left\vert \alpha_{2}(0)\right\vert ^{2}}\sum
_{n=0}^{\infty}\frac{\left(  \left\vert \alpha_{2}(0)\right\vert ^{2}\right)
^{n}}{n!}\gamma_{n}=\gamma_{eff}\label{taylor-04}%
\end{align}
then from (\ref{taylor-01}) and (\ref{taylor-04}), we have%
\begin{equation}
\left\vert \rho_{12}^{(ND)}(t)\right\vert \cong\left\vert ab\right\vert
\exp\left(  -\gamma_{eff}t\right) \label{taylor-05}%
\end{equation}
This is precisely the interpolation that corresponds to eq. (\ref{ft}). Now we
have the decoherence time
\begin{equation}
t_{D}=\frac{1}{\gamma_{eff}}=\frac{2}{m\omega}\frac{1}{L_{0}^{2}}%
t_{R}\label{44'}%
\end{equation}
In fact this $t_{D}^{-1}$ turns out to be a weighted average of the imaginary
part of the poles $z_{n}$. The same time was found by Omn\`{e}s in
\cite{Omnesazul} (or in eq. (\ref{4-0-01}) of this paper) and corresponds to
the definition (\ref{34'}) of the last section. So in fact, we have found the
same result. Also in \cite{Omnesazul} the result for $t_{D}$ is only valid for
small $t$ as in the last section. So the coincidence of both formalisms is proved.

Let us now consider the mathematical definition of moving preferred basis. The
basis $\{|\alpha_{1}(t)\rangle,|\alpha_{2}(t)\rangle\}=\{|\alpha_{i}%
(t)\rangle\}$, is orthonormal when $L\rightarrow\infty$ and the reasoning
below is done under this condition. Then let us diagonalize $\rho_{P}(0)$
(always when $L\rightarrow\infty$) as%
\begin{equation}
\rho_{P}(t)=\sum_{i=1,2}\rho_{i}(t)|\widetilde{i(t)}\rangle\langle
\widetilde{i(t)}|=\rho_{1}(t)|\widetilde{1(t)}\rangle\langle\widetilde
{1(t)}|+\rho_{2}(t)|\widetilde{2(t)}\rangle\langle\widetilde{2(t)}|
\end{equation}
where $\{|\widetilde{i(t)\rangle}\}$ is our orthogonal moving pointer basis.
But for times $t\gg t_{D}$ $\rho_{R}(t)=\rho(t)^{(D)}$ we have%
\begin{equation}
\rho_{P}(t)=\rho^{(D)}(t)=\left\vert a\right\vert ^{2}\left\vert \alpha
_{1}(t)\right\rangle \left\langle \alpha_{1}(t)\right\vert +\left\vert
b\right\vert ^{2}\left\vert \alpha_{2}(t)\right\rangle \left\langle \alpha
_{2}(t)\right\vert
\end{equation}
Then, since a linear orthonormal decomposition is unique we find the moving
pointer basis
\begin{equation}
|\widetilde{1(t)}\rangle=\left\vert \alpha_{1}(t)\right\rangle ,\text{
\ }|\widetilde{2(t)}\rangle=\left\vert \alpha_{2}(t)\right\rangle
\end{equation}
Then Omn\`{e}s basis coincides with $\{|\widetilde{1(t)}\rangle,|\widetilde
{2(t)}\rangle\}$, but this also is our moving preferred basis since it evolves
under the slow motion pole evolution. So Omn\`{e}s basis and our basis
coincide (always when $L\rightarrow\infty$).

So we have proved that all the characters of the Omn\`{e}s model: $t_{R},$
$t_{D}$, and the moving preferred basis, coincides with our definitions of the last\ section.

\section{\label{Conclusions}Conclusions}

In this work we have:

i.- Discussed a general scheme for decoherence, that in principle could be
used in many examples.

ii.- We have given a quite general definition of moving preferred basis
$\{\widetilde{|j(t)\rangle}\}$, and of relaxation and decoherence times for a
generic system.

iii.- We have proved that our definitions coincide with those of the Omn\`{e}s model.

We hope that these general results will produce some light in the general
problem of decoherence.

The Omn\`{e}s formalism, of references \cite{OmnesPh}, \cite{OmnesRojo}, and
\cite{Omnesazul} contain the most general definition of moving preferred basis
of the literature on the subject. Our basis has another conceptual frame: the
catalogue of decaying modes in the non-unitary evolution of a quantum system.
But since the Omn\`{e}s formalism is the best available it would be very
important for us to show, in the future, the coincidence of both formalisms,
as we have at least done for one model in this paper.

Of course we are fully aware that, to prove our proposal, more examples must
be added, as we will do elsewhere. But we also believe that we have a good
point of depart. In fact, probably the coincidences that we have found in the
Omn\`{e}s model could be a general feature of the decoherence phenomenon.
Essentially because, being the poles catalogue the one that contains
\textit{all the possible decaying modes} of the non unitary evolutions, since
relaxation and decoherence are non-unitary evolutions, necessarily they must
be contained in this catalogue.

\begin{center}
\emph{Acknowledgments}
\end{center}

We are very grateful to Roberto Laura, Olimpia Lombardi, Roland Omn\`{e}s and
Maximilian Schlosshauer for many comments and criticisms. This research was
partially supported by grants of the University of Buenos Aires, the CONICET,
the FONCYT, the UCEL of Argentina, and the Max-Planck-Institut f\"{u}r Physik
komplexer Systeme of Germany.

\end{document}